\documentclass[12pt]{article}
\usepackage{slashbox}
\usepackage{multirow}
\usepackage[T1]{fontenc}
\usepackage{algpseudocode}
\usepackage{algorithm}

\usepackage{graphicx}
\usepackage{enumerate}
\usepackage{rotating}
\usepackage{tikz}
\newcommand{\myComment}[3]{% szerokosc odstep text
   \hskip #2 \parbox[t]{#1}{\Comment{\noindent #3}}
}

\usepackage{amsmath, amsthm, amsfonts, amssymb}

\newtheorem{theorem}{Theorem}
\newtheorem{lemma}{Lemma}
\newtheorem{corollary}[theorem]{Corollary}

\newtheorem{proposition}[theorem]{Proposition}
\newtheorem{definition}{Definition}

\newenvironment{prooftw}[2][Proof of Theorem]
{\par\noindent{\bf #1 #2.} }{\hspace*{\fill}\nolinebreak{$\Box$}\bigskip\par}
\newenvironment{prooff}[1][Proof]
{\par\noindent{\bf #1.} }{\hspace*{\fill}\nolinebreak{$\Box$}\bigskip\par}

\begin{document}
%\markboth{}{}
\title{\bf On bipartization of cubic graphs by removal of an independent set\footnote{This project has been 
partially supported by Narodowe Centrum Nauki under contract 
DEC-2011/02/A/ST6/00201}}
\date{}
\author{Hanna Furma\'nczyk\footnote{Institute of Informatics,\ University of Gda\'nsk,\ Wita Stwosza 57, \ 80-952 Gda\'nsk, \ Poland. \ e-mail: hanna@inf.ug.edu.pl},  \ Marek 
Kubale\footnote{Department of Algorithms and System Modelling,\ Technical University of Gda\'nsk,\ Narutowicza 11/12, \ 80-233 Gda\'nsk, \ Poland. \ e-mail: kubale@eti.pg.gda.pl}, 
Stanis\l{}aw Radziszowski\footnote{Department of Computer Science,\ Rochester Institute of Technology,\ Rochester,\ NY\ 14623, US. \ e-mail:  spr@cs.rit.edu}}

%\markboth{H. Furmańczyk, M. Kubale}{Some remarks on unequitable coloring of cubic graphs}

\maketitle
\begin{abstract}
We study a new problem for cubic graphs: bipartization of a cubic graph $Q$ by deleting sufficiently large independent set $I$. 
It can be expressed as follows: \emph{Given a connected $n$-vertex tripartite cubic graph $Q=(V,E)$ with independence number $\alpha(Q)$, 
does $Q$ contain an independent set $I$ of size $k$ such that  $Q-I$ is bipartite?} We are interested for which value of $k$ the answer to this question is affirmative.
We prove constructively that if $\alpha(Q) \geq 4n/10$, then the answer is positive for each $k$ fulfilling
$\lfloor (n-\alpha(Q))/2 \rfloor \leq k \leq \alpha(Q)$. It remains an open question if a similar construction is possible for cubic graphs with $\alpha(Q)<4n/10$.
  
Next, we show that this problem with $\alpha(Q)\geq 4n/10$ and $k$ fulfilling inequalities $\lfloor n/3 \rfloor \leq k \leq \alpha(Q)$ can be related to semi-equitable graph 3-coloring, 
where one color class is of size $k$, and the subgraph induced by the remaining vertices is equitably 2-colored. This means that $Q$ has a coloring of type 
$(k, \lceil(n-k)/2\rceil, \lfloor (n-k)/2 \rfloor)$. 

\end{abstract}

\section{Some preliminaries}

There are many challenging and interesting problems involving independent sets and cubic graphs. One of the most known is the problem of independence, IS$(Q,k)$:

\begin{quote}
Given a connected cubic graph $Q = (V,E)$ and integer $k$, does $Q$ contain an independent set of size at least $k$?
\end{quote}

An \emph{independent set} of a graph $Q$ is a subset of the vertices of $Q$ such that no two vertices in the subset are joined by an edge in $Q$. 
The size of the largest independent set is called the \emph{independence number} of $Q$, and it is denoted by $\alpha(Q)$. The problem of finding the value of $\alpha(Q)$ is 
also widely discussed in the literature.
In general, the problem IS$(Q,k)$ is NP-complete for cubic graphs, and even for planar cubic graphs \cite{npis}. %Heckman et al. \cite{heckman} proved that every triangle-free subcubic planar graph has an independent set with size at least $0.375 n$. 
A comprehensive survey of results on the IS problem, including cubic graphs, was presented in \cite{fastIS, heckman}.

The second type of problems is connected with decycling sets of cubic graphs (also known as feedback-vertex sets). For a graph $Q$, a subset $S\subseteq V(Q)$ is a 
\emph{decycling set} of $Q$ if and only if $Q - S$ is acyclic, where by $Q-S$ we mean the subgraph of $Q$ induced by the vertices in $\overline{S}=V(Q) \setminus S$. The cardinality of a 
smallest decycling set of $Q$ is called the \emph{decycling number}, and it is denoted by $\Phi(Q)$. Speckenmeyer \cite{feed} showed that cubic graph $Q$ has $\Phi(Q) = n/2 - z(Q) +1$, where $z(Q)$ is the size of maximum nonseparating independent set $J$ ($Q-J$ must be connected). 
Recently, Nedela and Kotrbcik \cite{nedela} expressed the value of $\Phi(Q)$ in terms of odd components in $Q-E(T)$: 
\begin{equation}
\Phi(Q)=\frac{n}{4} + \frac{\xi(Q)+1}{2}, 
\end{equation}
where $Q$ is a cubic graph, and $\xi(Q)$ denotes the minimum number of odd components in $Q-E(T)$ over all spanning trees $T$. Speckenmeyer \cite{feed} obtained an upper bound on 
$\Phi(Q)$ in terms of the girth $g$ of a cubic graph $Q$:
\begin{equation}
\Phi(Q)\leq \frac{g+1}{4g-2}n + \frac{g-1}{2g-1}.\label{speck}
\end{equation}
For $g=3$, the bound (\ref{speck}) gives $\Phi(Q) \leq 4n/10 + 4/10$. The bound (\ref{speck}) was improved by Liu and Zhao \cite{liu} for most cubic graphs, except $K_4$, two other cubic 
graphs and one subclass defined by the authors, to: 
\begin{equation}
\Phi(Q) \leq 
\frac{g}{4(g-1)}n+\frac{g-3}{2g-2}.\label{drugie}
\end{equation}
 For $g=3$, the bound (\ref{drugie}) gives $\Phi(Q) \leq 3n/8$.

The third group contains problems connected with \emph{bipartization} of cubic graphs. Given a graph, the task is to find a smallest set of vertices whose deletion makes the remaining graph 
bipartite. Choi et al. \cite{bip} showed that the bipartization decision problem is NP-complete for cubic graphs. Some approximation algorithms were given in \cite{falk}.

We combine the above approaches and define the Bipartization IS, BIS$(Q,k)$ problem, as follows:
\begin{quote}
Given a connected cubic graph $Q=(V,E)$ and integer $k$, does $Q$ contain an independent set $I$ of size at least $k$ such that  $Q-I$ is bipartite?
\end{quote}
This problem can be seen as a task of finding independent odd decycling sets.

In the sequel we consider connected cubic graphs $Q$ with chromatic number $\chi(Q)=3$. This means that the set of vertices of graph $Q$ can be 
partitioned into three independent sets and $Q$ is not bipartite. The class of such cubic graphs will be denoted by $\mathcal{Q}_3$. Its subclass of 
graphs on $n$ vertices will be denoted by $\mathcal{Q}_3(n)$. Let us recall the Brooks theorem:

\begin{theorem}[\cite{brooks}]
For any connected undirected graph G with maximum degree $\Delta$, the chromatic number $\chi(G)$ of $G$ is at most $\Delta$, unless $G$ is a clique or an odd cycle.
\end{theorem} 

This implies that $$2\leq \chi(Q) \leq 3$$ for all cubic graphs except $K_4$. 

A graph is \emph{equitably $t$-colorable} if and only if its vertex set can be partitioned into independent sets $V_1,V_2,\ldots,V_t$ such that $\big||V_i|-|V_j|\big| \leq 1$ for all
 $i,j=1,2,\ldots,t$. Chen et al. \cite{chen} proved constructively that every cubic graph can be equitably colored without 
adding new colors. Hence, in particular for $Q \in \mathcal{Q}_3$ we have 
\begin{equation}
\chi_=(Q)=\chi(Q),\label{chen}
\end{equation}
where $\chi_=(G)$ is the equitable chromatic number. Their algorithm relies on descreasing the width of coloring (the difference between the cardinality of the largest and smallest 
color class) by one until the difference is not greater than one. 

In this paper we are also interested in equitable coloring of $Q-I$. We will give an algorithm which, given an independent set of size $k \geq 4n/10$, constructs an appropriate 
independent set $I$ of size $k$ for the BIS$(Q,k)$ problem 
with $Q \in \mathcal{Q}_3(n)$. We will also prove that such cubic graphs have colorings of type 
$(k, \lceil (n-k)/2\rceil, \lfloor (n-k)/2 \rfloor)$, which means that $Q-I$ has an equitable 2-coloring. Such type of coloring is called
\emph{semi-equitable}, i.e. the coloring in which exactly one color class is of any size while the cardinalities of the remaining color classes differ by at most 1. 
Colorings of this kind may be used in a problem of mutual exclusion scheduling of jobs on three uniform parallel processors \cite{baker}. In such a model of scheduling one of processors is 
faster than the remaining two, while the two 
slower processors are 
of the same speed and the conflict graph is cubic. 
%The problem is NP-hard for general cubic graphs, but in some special cases our algorithm may help to solve it in polynomial time.

\section{Main results}  

Our main result is as follows.
\begin{theorem}
If $Q\in\mathcal{Q}_3(n)$ and $\alpha(Q) \geq 4n/10$, then there exists an independent set $I$ of size $k$ in $Q$ such that $Q-I$ is bipartite for $\lfloor n/3 
\rfloor \leq k \leq \alpha(Q)$.\label{tw_res}
\end{theorem}
Before we prove Theorem \ref{tw_res}, we need some auxiliary concepts.

We consider connected cubic graphs $Q\in \mathcal{Q}_3(n)$ with independence number $\alpha(Q) \geq 4n/10$, and let $I$ be an independent set of size at 
least $4n/10$.
%($n$ is divisible by 10). 
If $Q-I$ is not bipartite then the subgraph $Q - I$ consists of two parts: a 2-chromatic part of all bipartite components and a 3-chromatic part containing odd cycles 
(possibly with chords, bridges, pendant edges, etc.).

\begin{definition}
\emph{For $Q\in\mathcal{Q}_3$, the} residuum $R(I)$ of $Q$ with respect to an independent set $I$ \emph{is the set of all odd cycles in the graph $Q-I$.}
\end{definition}

\begin{figure}
\begin{center}
\includegraphics[scale=0.9]{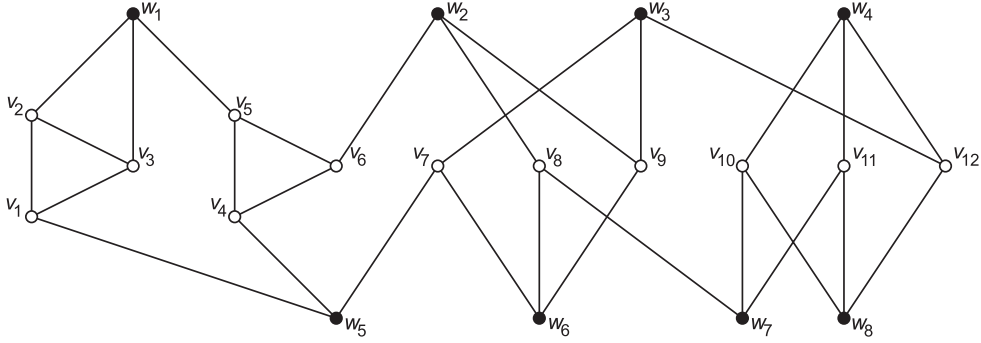} 
%\vspace{4cm}
\caption{\small{Example of a cubic graph in $\mathcal{Q}_3(20)$ with independent set $I$ of size 8. The vertices of $I$ are marked in black. $R(I)=\{v_1v_2v_3,v_4v_5v_6\}$.}}\label{res}
\end{center}
\end{figure}

For example, for the graph in Fig.~\ref{res} and given $I$, $R(I)=\{v_1v_2v_3,v_4v_5v_6\}$.

\begin{definition}
\emph{Vertex $w\in I$ is a} free vertex \emph{in $Q\in\mathcal{Q}_3$ with respect to independent set $I$ if and only if its removal from $I$ (but not from $V(Q)$) results in the same 
residuum, i.e.
$R(I) = R(I\setminus \{w\})$. The set of all free vertices in $I$ will be denoted by $F_0$.}
\end{definition}

For the graph in Fig.~\ref{res} and given independent set $I$, $F_0=\{w_2,w_3,\ldots, w_8\}$. Vertex $w_1$ is not free because moving it from $I$ to $Q-I$ creates a new odd cycle
in $Q-I$, namely $w_1v_2v_3$. Clearly, $F_0 \subset  I$.

\begin{definition}
\emph{A} diamond \emph{in $Q$ with respect to independent set $I$ is a subgraph $D$ on vertices $\{u,w,a,b\} \subseteq V(Q)$ isomorphic to $K_4-e$, where $u,w \in I$.}\end{definition}

\begin{figure}[htb]
\begin{center}
\includegraphics[scale=0.9]{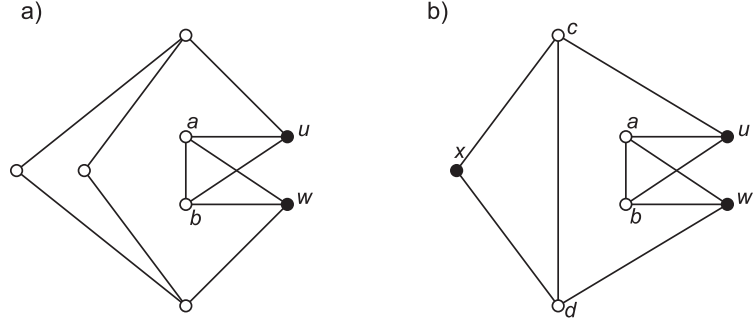} 
%\vspace{-4cm}
\caption{\small{Subgraph of $Q$ containing: a) diamond $K_4-e$ with pseudo-free vertices $u$ and $w$ of type 1; $u,w \in I$; b) diamond $K_4-e$ with pseudo-free vertices $u$ and $w$ of type 2; 
$u,w,x \in I$.}}\label{k4-e}
\end{center}
\end{figure}

\begin{definition}
\emph{Vertices $u, w \in I$ are} pseudo-free vertices of type $1$ \emph{in $Q$ with respect to independent set $I$ if and only if there is a diamond $D$ in $Q$ on vertices $\{u,w,a,b\}$, 
and there is no odd 
cycle $C$ of length at least 5 with vertices in $\overline{I} \cup \{u,w\}$ such that $|V(C) \cap V(D)|=3$, (cf. Fig. \ref{k4-e}a).} \emph{The set of all pseudo-free vertices of type 1 will be denoted 
by $F_1$.}\label{def1}
\end{definition}

Note that  $F_1 \subset  I$ and $F_1$ is a disjoint union of pairs of vertices $\{u,w\}$ satisfying Definition \ref{def1}.

\begin{definition}
\emph{Vertices $u, w \in I$ are} pseudo-free vertices of type $2$ \emph{in $Q$ with respect to independent set $I$ if and only if there is a diamond $D$ in $Q$ on vertices $\{u,w,a,b\}$, and 
there is a cycle $C_5$ with vertices in $\overline{I} \cup \{u,w\}$ such that $|V(C_5) \cap V(D)|=3$, and the two vertices $\{c,d\}=V(C_5) \setminus V(D)$ have a common neighbor $x$ in 
$I$ (cf. Fig. \ref{k4-e}b).} \emph{The set of all pseudo-free vertices of type 2 will be denoted by $F_2$.}\label{def2}
\end{definition}

Note that vertices $c$ and $d$ are consecutive on $C_5$, and $F_2 \subset  I$ is a disjoint union of pairs of vertices $\{u,w\}$ satisfying Definition \ref{def2}.

Let $F(I)=F_0 \cup F_1 \cup F_2$. Clearly, $F_i \cap F_j = \emptyset$ for $i,j=0,1,2$ and $i \neq j$.  

The following main auxilliary lemma implies, that under the assumptions of Theorem \ref{tw_res}, if  $R(I)$ is nonempty then so is $F(I)$.
\begin{lemma}
If $Q \in\mathcal{Q}_3(n)$ has an independent set $I$ of size at least $4n/10$ and $R(I) \neq \emptyset$, then there exists a free or pseudo-free 
\emph{(}of type \emph{1} or \emph{2)} vertex in $I$. 
\label{lmnon-emptyF}
\end{lemma}
We will prove Lemma \ref{lmnon-emptyF} in Section \ref{dowod}.

\begin{prooftw}{\ref{tw_res}}
First, we will prove that our theorem holds for $k$ fulfilling $4n/10 \leq k \leq \alpha(Q)$.

Let $I$ be any independent set of size at least $4n/10$. Assume that $R(I) \neq \emptyset$. We will show that there exists another independent set, say $J$, such that $|J| \geq
|I|$ and $R(J) \subsetneq R(I)$.

Let $C$ be an odd cycle belonging to $R(I)$. Any vertex $v \in V(C)$ must be of degree 2 or 3 in $Q-I$. 
If there exists $v\in V(C)$ of degree 3 in $Q-I$, then we set $J= I \cup \{v\}$. The new residuum $R(J)$ is a subset of $R(I)\setminus \{C\}$. 
Otherwise, if each $v \in V(C)$ is of degree 2 in $Q-I$, then let $v_1\in V(C)$ and $P=v_1 v_2 \ldots v_k$ be the shortest path from $v_1$ to $v_k \in F(I)$ in $Q$ such that none of vertices $v_1, v_2, \ldots, v_{k-1}$ is free or 
pseudo-free. We know from Lemma \ref{lmnon-emptyF} that $F(I)$ is nonempty.
We consider two cases.

\begin{description}
\item[\textnormal{Case 1:}] \emph{$P$ is a path alternating between $I$ and $\overline{I}$.}

This means that $v_1,v_3,\ldots, v_{k-1} \in \overline{I}$ and $v_2,v_4, \ldots, v_k \in I$, $v_k=w$.
\begin{description}
\item[\textnormal{Subcase 1.1:}] \emph{Each vertex of $v_3,v_5,\ldots, v_{k-1}$ has exactly two neighbors in $I$.}

Then we interchange even and odd vertices between $I$ and $\overline{I}$ along the path $P$ so that a new independent set
$$J =\left\{
\begin{array}{ll}
I \cup \{v_1, v_3,\ldots,v_{k-1}\} \setminus \{v_2,v_4,\ldots, v_k\} & \text{if } w \in F_0,\\
I \cup \{v_1, v_3,\ldots,v_{k-1},a\} \setminus \{v_2,v_4,\ldots, v_k,u\}  & \text{if } w \in F_1 \cup F_2, \end{array}
\right.
$$
of the same size is obtained, and $R(J) \subset R(I)\setminus \{C\}$. 
\item[\textnormal{Subcase 1.2:}] \emph{There is a vertex in $\overline{I}$ on path $P$ such that all \emph{(}three\emph{)} of its neighbors belong to $I$.}

In this case we choose among such vertices one with the smallest index, say $v_i$, $3\leq i \leq k-1$ (see vertex $v_5$ in Fig. \ref{subcase1_2}). Let us observe that vertex $v_{i-1} \in I$ can belong 
to odd cycles 
in $Q - (I \setminus \{v_{i-1}\})$, including the edge $\{v_{i-2}, v_{i-1}\}$, but there is no odd cycle in $Q - (I \setminus \{v_{i-1}\})$ passing through $\{v_{i-1}, v_i\}$. We interchange 
even and odd vertices along the prefix subpath $v_1 v_2 \ldots v_{i-1}$ of $P$, so that $J=I \cup \{v_1, v_3,\ldots,v_{i-2}\} \setminus \{v_2,v_4,\ldots, v_{i-1}\}$, and
$R(J) \subset R(I)\setminus \{C\}$.

\begin{figure}[htb]
\begin{center}
\includegraphics{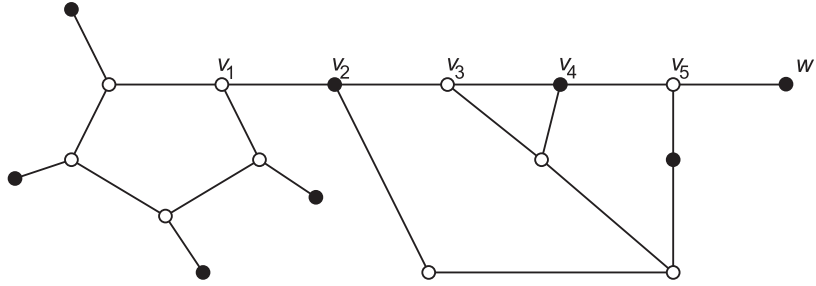} 
%\vspace{-4cm}
\caption{\small{Example of a subgraph of $Q$ with alternating path $P=v_1v_2v_3v_4v_5w$, where vertex $v_5$ is as $v_i$ of Subcase 1.2.}} \label{subcase1_2}
\end{center}
\end{figure}
\end{description}
\item[\textnormal{Case 2:}]  \emph{$P$ is not an alternating path.}

This means that there is a vertex in $Q - I$ on $P$ such that its successor on path $P$ is also in $\overline{I}$. We choose among such vertices one with the smallest index, 
say $v_j$ (see vertex $v_5$ in Fig. \ref{case2}). 
\begin{figure}[htb]
\begin{center}
\includegraphics{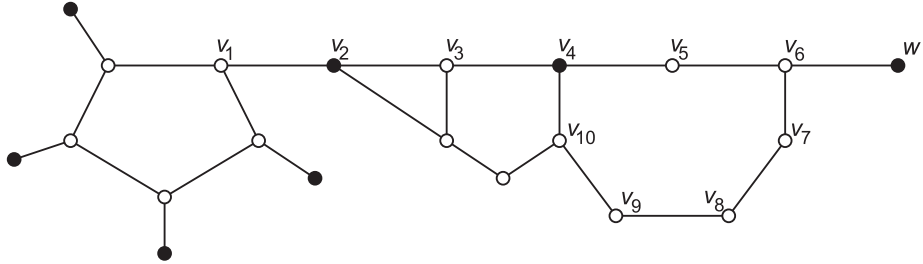} 
%\vspace{-4cm}
\caption{\small{Example of a subgraph of $Q$ and path $P=v_1v_2v_3v_4 v_5v_6w$, which is not alternating. Vertices $v_2, v_4, w \in I$, vertex $v_5$ is as $v_j$ in Case 2. After 
applying
the procedure described in Case 2: $v_1, v_3 \in I'$ while $v_2,v_4 \in \overline{I'}$ and there is a cycle $C'=v_4v_5v_6v_7v_8v_9v_{10}$ with $v_t=v_6$ (Subcase 2.2).}} \label{case2}
\end{center}
\end{figure}

We consider the alternating part of the path $P$ up to vertex $v_{j-1}$ as in Case 1. Let $I'$ be a new independent set obtained after applying the procedure from Case 1. 
Now, we have one of the following subcases:
\begin{description}
\item[\textnormal{Subcase 2.1:}] \emph{There is no odd cycle in $Q - I'$ containing the edge $\{v_{j-1},v_j\}$.}
 
Cycle $C$ is broken and no new odd cycle is created. $J = I'$.
\item[\textnormal{Subcase 2.2:}] \emph{There is an odd cycle $C'$ in $Q - I'$ containing the edge $\{v_{j-1},v_j\}$.} 

If there exists $v\in V(C')$ of degree 3 in $Q-I'$, then we set $J= I' \cup \{v\}$ (the new residuum $R(J)$ is a subset of $R(I')\setminus \{C'\}$; of course, $R(J) \subsetneq R(I)$).  
Otherwise, let $v_t$ be the vertex belonging to both $C'$ and $P$ whose index $t$ is the highest (see vertex $v_6$ in Fig. \ref{case2}). Note, that $v_{t+1} \in I'$. In this case we 
consider the alternating part of the path $P$ starting with vertex $v_t$ as in Case 1, and finally obtain $J$, which clearly satisfies $R(J) \subsetneq R(I)$.
\end{description}
\end{description}

If the new independent set has still a nonempty residuum, we repeat our algorithm iteratively (with another cycle $C$ and path $P$). There is at least one odd cycle broken in each iteration of 
the algorithm. Consequently, after $s$ iterations of the algorithm, we obtain a 
sequence of independent sets $J_1,J_2, \ldots, J_s$ of non-decreasing sizes, $R(J_s)=\emptyset$, and hence $Q - J_s$ is bipartite. 

Therefore, by Lemma \ref{lmnon-emptyF}, if a cubic graph $Q\in \mathcal{Q}_3$ has an independent set $I$ of size $k \geq 4n/10$, then it also has an independent set $J_s$ of size at least $k$ such that 
$Q-J_s$ is bipartite. Due to Chen's constructive proof thus obtained 3-coloring of $Q$
can be equitalized to $(\lfloor n/3 \rfloor, \lfloor (n+1)/3 \rfloor, \lfloor (n+2)/3 \rfloor)$ by decreasing the width of the coloring one by one, which completes
the proof for all $k$, 
$\lfloor n/3 \rfloor \leq k \leq \alpha(Q)$. 

%This means that one of middle 
%colorings implies independent set $I''$ of size exactly $k=|I|$ such that $Q-I''$ is bipartite. 
\end{prooftw}

Now, we summarize in pseudocode the procedures of our construction. For each iteration of the algorithm the path $P=v_1 \ldots v_k$
in given cubic graph $Q$ is fixed, similarly as $Q$. Indices $in$ and $out$ refer to 
the segment of the path $v_{in} \ldots v_{out}$.

\begin{algorithm}
\caption{Decycling Algorithm}
\begin{algorithmic}
\Require {Cubic graph $Q$ with independent set $I$ of size $k$, $4n/10 \leq k \leq \alpha(Q)$, and an odd cycle $C \in R(I)$.}
\Ensure {Independent set $J$ of size at least $k$ such that $R(J) \subsetneq R(I)$.}
%\Procedure {Decycling}{$Q$,$I$}
%\State $v \leftarrow$ vertex on $C$ of degree 3 in $Q - I$ \myComment{2in}{0.7in}{$v = 0$ if such vertex does not exist}
\If {there exists a vertex $v$ on $C$ of degree $3$ in $Q-I$} \Return $I \cup \{v\}$ \EndIf
\State $P \leftarrow v_1\ldots v_k$ \myComment{4in}{0.7in}{$v_1$ is on cycle $C$, $v_k \in F(I)$, and $v_j \notin F(I)$ for $2 \leq j \leq k-1$}
\State $in \leftarrow 1$
%\Repeat
\State $out \leftarrow$ \Call{AlternatingPath}{{$P, in$}} \myComment{2.5in}{0.7in}{segment of $P$ from $v_{in}$ to $v_{out}$ is alternating as in Case 1}
\State $J \leftarrow$ \Call{NewIndependentSet}{{$P, in, out, I$}}
%\State $J \leftarrow I'$
\While {$|R(I)| = |R(J)|$}
%\If {$R(I') \subset R(J)$} Return $I'$ \EndIf
\State $I \leftarrow J$ 
%\State $J \leftarrow I'$
\State $in \leftarrow out$
\State $(out,J) \leftarrow$ \Call{NonAlternatingPath}{{$P, in, I$}} \myComment{1.65in}{0.7in}{segment of $P$ from $v_{in}$ to $v_{out}$ is non-alternating as in Case 2}
%\If {$R(I') \subset R(I)$} \Return $I'$ \EndIf
%\State $I \leftarrow I'$ \State $in \leftarrow out$ \State \textbf{goto} 3
\EndWhile
%\Until $R(I')=\emptyset$
\State \Return $J$
%\EndProcedure
\end{algorithmic}
\end{algorithm}

\floatname{algorithm}{Procedure}

\begin{algorithm}
\caption{Alternating Path}
\begin{algorithmic}
\Require {Path $P=v_1 \ldots v_k$, integer $in$.}
\Ensure {Integer $out$ such that $v_{out} \in \overline{I}$ is a vertex with the smallest index on the alternating part of path $P$: 
$v_{in}\ldots v_{out}$,
such that all its neighbors are from $I$ if such vertex exists or, otherwise, $v_{out} \in \overline{I}$ 
is the last vertex on alternating part of the path $P$.}
\Procedure {AlternatingPath}{$P,in$}
\State $i \leftarrow in$ \myComment{3in}{2in}{$v_i$ is starting point of the alternating part of the path $P$}
\Repeat
\If {$i < k$} $i \leftarrow i+1$ \EndIf
\If {$v_i \in Q - I$} \Return $i-1$ \EndIf \myComment{3in}{2in}{two consecutive vertices on path $P$
are from $\overline{I}$, vertex $v_{i-1} \in \overline{I}$ is the last
vertex on alternating part of $P$ started in $v_{in}$}
\If {$i=k$} \Return $i$ \EndIf \myComment{3in}{2in}{an alternating part of $P$, started in $v_{in}$, ends in $v_{k}$} 
\State $i \leftarrow i+1$ 
\Until {$v_i \in \overline{I}$ is a vertex whose all neighbors belong to $I$} 
\State \Return $i$ 
\EndProcedure
\end{algorithmic}
\end{algorithm}

\begin{algorithm}
\caption{New Independent Set}
\begin{algorithmic}
\Require {Path $P=v_1\ldots v_k$, integers $in$ and $out$ delimiting alternating part of $P$, independent set $I$.}
\Ensure {Independent set $I'$.}
\Procedure {NewIndependentSet}{$P, in, out, I$} \Comment{Case 1}
\If {$out < k+1$ \textbf{or} ($out=k+1$ \textbf{and} $v_k \in F_0$)} 
\State $I' \leftarrow I \cup \{v_{in}, v_{in+2},\ldots, v_{out-2}\}\setminus \{v_{in+1}, v_{in+3},\ldots, v_{out-1}\}$ \Else
%\If {$out = k+1$ \textbf{and} $v_k=w \in F_1 \cup F_2$} 
\State $I' \leftarrow I \cup \{v_{in}, v_{in+2},\ldots, v_{out-2},a\}\setminus \{v_{in+1}, v_{in+3},\ldots, v_{out-1}, u\}$ \EndIf
\State \Return $I'$
\EndProcedure
\end{algorithmic}
\end{algorithm}

\begin{algorithm}
\caption{Non-alternating Path}
\begin{algorithmic}
\Require {Path $P=v_1\ldots v_k$, integer $in$ such that $v_{in}$ is a starting point of non-alternating part of the path $P$, independent set $I$.}
\Ensure {Integer $out$, independent set $J$.}
\Procedure {NonAlternatingPath}{$P, in, I$}
\State $C' \leftarrow$ odd cycle in $R(I)$ with edge $\{v_{in-1},v_{in}\}$ %\myComment{1.4in}{0.7in}{$C' = \emptyset$ if such cycle does not exist}
%\State $v \leftarrow$ vertex on $C'$ of degree 3 in $Q - I$ \myComment{2in}{0.7in}{$v = 0$ if such vertex does not exist}
\If {$C'$ has a vertex $v$ of degree $3$ in $Q-I$} \Return $(k,I \cup \{v\})$ \EndIf

\State $i \leftarrow \min\limits _{j > in} \{j | v_j \notin V(C')\}$
%\Repeat \State $i \leftarrow i+1$  \Until{$v_i$ is not a vertex of $C'$} 
\If {$v_i\in I$} 
\State $out \leftarrow i-1$ 
\State $J \leftarrow I$ \Else
%\State $v_t \leftarrow v_{i-1}$
\State Set $J$ as in Subcase 2.2 with $t=i-1$
\State $out \leftarrow i$ \EndIf
\State \Return $(out, J)$
\EndProcedure
\end{algorithmic}
\end{algorithm}

\newpage
\section{Proof of Lemma \ref{lmnon-emptyF}} \label{dowod}
\setcounter{lemma}{0}
\begin{lemma}
If $Q \in\mathcal{Q}_3(n)$ has an independent set $I$ of size at least $4n/10$ and $R(I) \neq \emptyset$, then there exists at least one free or pseudo-free 
\emph{(}of type \emph{1} or \emph{2)} vertex in $I$.
\label{lmnon-emptyF}
\end{lemma}
\begin{prooff}
We need to prove that $F(I)=F_0\cup F_1 \cup F_2 \neq \emptyset$. 
First, we assume that $10 | n$ and let $I$ be an independent set of size $4n/10$. Let $L$ denote the set of isolated 
vertices in $Q-I$, $|L|=l$. We are interested in the structure of $Q-I$, including the value of $l$. This is a graph 
with $6n/10$ vertices and $3n/10$ edges. Let us notice that if $Q-I$ has no isolated vertices, 
$Q-I$ must define a perfect matching, in which case we have $3n/10$ $K_2$'s. If $Q-I$ contains components with more than one edge, then we have some number of isolated vertices. For example, a cycle 
$C_k$ in $Q-I$ ''implies'' $k$ isolated vertices, and a path $P_k$ ''implies'' $k-2$ vertices. 
In general, a component of $Q-I$ with $m'$ edges and $n'$ vertices ''implies'' $2m'-n'$ isolated vertices. In most cases, $Q-I$ contains some components $K_1$, $K_2$, 
$P_k$, $C_{k'}$ ($k,k' \geq 3$), and a subgraph whose 
components have at least one vertex of degree 3. Let $Q_l$ denote the part of $Q-I$ excluding $K_1$'s and $K_2$'s, i.e. the part which ''implies'' the isolated vertices. For given $Q$ and $I$, 
the subgraph $Q-I$ consists of $Q_l$, $lK_1$ and $k_2K_2$.

We fix number $l$. In the following we consider two cases depending on its value.
\begin{description}
\item[\textnormal{Case 1:}] $l > 4n/30$.

Since $3l >|I|$, there must exist at least one vertex in $I$, say $u$, which is adjacent to at least two vertices in $L$. Note that 
$u$ cannot be on any cycle together with vertices from $Q-I$. This means that $u \in F_0$ is a free vertex and $F_0 \neq \emptyset$. 

\item[\textnormal{Case 2:}] $l \leq  4n/30$.

In this case $F_0$ may be empty. We will show that if $F_0=F_1=\emptyset$, then $F_2\neq \emptyset.$

First, we will prove that there exists $K_2$ among $k_2K_2$'s such that it is a subgraph of a diamond ($K_4-e$). 
We introduce some additional
notation. Let $\mathcal{K}^i$ denote the set of all such $K_2$'s in $Q-I$, whose endvertices have exactly $i$ common neighbors in $I$, and let $|\mathcal{K}^i|=k_2^i$,
$i=0,1,2$. Of course, $k_2=k_2^0+k_2^1+k_2^2$. Moreover, let us notice that $k_2^2K_2$'s result in $k_2^2$ diamonds.

\noindent\textit{Claim} There is a diamond in $Q$, i.e. $k_2^2 >0$.

\noindent\textit{Proof of Claim}.
\\For a contradiction, let us assume that the endvertices of each $K_2$ have at most one common neighbor in $I$.

Now, we ask how many non-free vertices in $I$ might be implied by vertices from $Q-I$ assuming also $F_0 =\emptyset$. 
Let $\gamma$ denote the number of non-free vertices in $I$ that are adjacent to vertices from $Q_l$ and are on some odd cycle (resulting from the fact that they are 
non-free vertices) on vertices from $Q_l \cup I$. We say that $Q_l$ ''generates'' $\gamma$ non-free vertices, $\gamma \leq \gamma_{\max},$ where $\gamma_{\max}$ is upper bound on $\gamma$
for fixed subgraph $Q_l$, excluding the structure of $Q \setminus Q_l$.

The number we asked for is equal to $\gamma_{max}+k_2$. 

We show that

\begin{equation}
3n/10-2l+3 \leq k_2 \leq 3n/10 -3l/4.\label{k2_bound}
\end{equation}
 
Indeed, let us consider the stucture of $Q_l$  implying the minimal number of $K_2$'s. It is easy 
to see that such $Q_l$ must 
contain $C_3$ and $(l-3)P_3$, with $3(l-2)$ vertices and $2(l-2)+1$ edges. This implies that $k_2 \geq (6n/10-l-3(l-2))/2=3n/10 - 2l +3$.
On the other hand, the structure of $Q_l$ maximizing $k_2$ must contain the minimal number of vertices equal to $6n/10-2k_2-l$ with $3n/10-k_2$ edges. Hence $k_2$ satisfies
$3(6n/10-2k_2-l) \geq 6n/10-2k_2$, which implies the upper bound in (\ref{k2_bound}).

%Let us notice that small value of $\gamma_{\max}$ is equivalent with large value of $k_2$ and vice versa.

The maximal value of $\gamma_{\max}$ is equal to $3l/2 - 7/2$ and it is achieved by $Q_l=C_3 \cup (l-3)/p\ P_{p+2}$ for even $p$. Let us notice that 
the number of vertices in such $Q_l$ is maximal for $p=2$. In this case $k_2= 3n/10 -3l/2+ 3/2$. We have

%$$\gamma_{\max} \leq \frac{3}{2} \ l -\frac{7}{2} \leq \frac{3}{2} * \frac{4}{30}n - \frac{7}{2} = \frac{2}{10} n-\frac{7}{2}.$$ 

\begin{equation}
\gamma_{\max}+k_2 \leq (3l/2-7/2)+(3n/10-3l/2+3/2) <4n/10.
\end{equation} 

On the other hand, for $Q-I$ with $R(I)\neq \emptyset$ the maximal number of $K_2$'s, due to inequalities (\ref{k2_bound}) does not exceed $3n/10-3l/4$. The value of $\gamma_{max}$ 
depends also on the 
structure of $Q_l$ with this number of $K_2$'s, but in any case $\gamma_{max} +k_2 < 4n/10$.
This means that $F_0 \neq \emptyset$, which is a contradiction.

\noindent\textit{End of proof of Claim.}

Now, let us assume that $F_1=F_0=\emptyset$. This means that each diamond $D$ from $\mathcal{K}^2$ is on an odd cycle 
of length at least 5. We note that such odd cycles can be caused by joining vertices from $I \cap V(D)$ to endvertices of $K_2$ from $\mathcal{K}^1 \cup \mathcal{K}^0$. 
Observe that diamonds connected in this way to $K_2 \in \mathcal{K}^1$ result in pseudofree 
vertices of type 2. 
%This means that $F_2\neq \emptyset$. 

Finally, assume that $F_2=\emptyset$. This implies that the endvertices of 
each $K_2\in \mathcal{K}^2$ are joined to vertices from $Q_l$ or 
to endvertices of $K_2\in \mathcal{K}^0$. Since $Q$ is cubic and connected, there is at most one diamond joined to each of $K_2 \in \mathcal{K}^0$.

Since $|V(Q_l)|+l+2k_2=6n/10$, and $k_2=k_2^0+k_2^1+k_2^2$, and by (\ref{k2_bound}), we have
$$|V(Q_l)|+l+k_2^0+k_2^1+k_2^2=6n/10 - k_2 \leq 3n/10 +2l -3.$$

Due to: $\gamma \leq |V(Q_l)|/2$, we get
$$2\gamma +l+k_2^0+k_2^1+k_2^2 \leq 3n/10 +2l -3.$$

Because clearly
$$
\gamma +2k_2^2 +k_2^1=4n/10=|I|,
$$

and hence $k_2^1 = 4n/10 -\gamma -2k_2^2-k_2^1$, we have:

\begin{equation}
k_2^2 \geq \gamma +n/10 +k_2^0+3 > \gamma +k_2^0.\label{contr}
\end{equation}

Since $F_1=\emptyset$, inequality (\ref{contr}) means that at least one diamond is joined to $K_2 \in \mathcal{K}^1$. This implies $F_2 \neq \emptyset$, a contradiction.
\end{description}
\end{prooff}

\section{Bipartization and equitable colorings}

In the paper we have posed a new problem for cubic graphs $Q \in \mathcal{Q}_3(n)$ with an independent set $I$ of size $k$. 
We answered the question about existence of appropriate bipartizing set for $\lfloor n/3 \rfloor \leq k \leq \alpha(Q)$, if only $\alpha(Q) \geq 4n/10$.

On the other hand, Frieze and Suen \cite{aprox} showed that the independence number of almost all cubic graphs on $n$ vertices 
satisfies $\alpha(Q) \geq 4.32n/10 - \epsilon n$, for any constant $\epsilon > 0$. Moreover, they gave a simple greedy algorithm which find an independent set of that size in almost all cubic
graphs. In practice this means that a graph from $\mathcal{Q}_3(n)$ is very likely to have an independent set 
of size $k \geq 4n/10$.

Taking into consideration the structure of the bipartized subgraph $Q-I$, it turns out that such a subgraph can be colored in equitable way with two colors. Let us assume that $|I|=4n/10$. 
Notice that 
$6n/10$ vertices of $Q-I$ induce binary trees (some of them may be trivial) and/or graphs whose 2-core is equibipartite (an even cycle possibly 
with chords). Note that deleting an independent set $I$ of cardinality $4n/10$ from a cubic graph $Q$ means also that we remove 
$12n/10$ edges from the set of all $15n/10$ edges of $Q$. The resulting graph $Q-I$ has $6n/10$ vertices and $3n/10$ edges. Let $s_i$, $0 \leq i \leq 3$, be 
the number of 
vertices 
in $Q-I$ of degree $i$, $\Sigma_{i=0}^3s_i=6n/10$. Since the number of edges is half the number of vertices, the number of isolated vertices, 
$s_0$, is 
equal to $s_2 + 2s_3$. If $s_0=0$, then $Q-I$ is a perfect matching and its equitable coloring is obvious. 

Suppose that $s_0 > 0$. Consider the part of $Q-I$ without isolated vertices and its 2-coloring. Each vertex of degree 3 causes the difference between cardinalities of color classes equal 
to at most 2 ($K_{1,3}$), similarly each vertex of degree 2 causes the difference at most 1 ($K_{1,2}$). The difference between the cardinalities of color classes in any 
coloring fulfilling these conditions is at most $s_2+2s_3$ in $Q-I-L$, and an appropriate assignment of 
colors to isolated vertices $L$ makes the graph $Q-I$ equitably 2-colored.
Hence, we have:

\begin{proposition}
If $Q\in\mathcal{Q}_3(n)$ has an independent set $I$ of size $|I| = 4n/10$ then it has a semi-equitable coloring of type $(4n/10, \lceil 3n/10 \rceil,\lfloor 3n/10 \rfloor)$.
\end{proposition}

Note, that if an $n$-vertex cubic graph $Q$ has an independent set $I$ of cardinality $|I| > 4n/10$ and consequently there exists independent set $J$ of the same cardinality 
such that $Q-J$ is bipartite, then we have more isolated vertices in $Q-J$ and a partition of $Q-J$ into $V_1$ and $V_2$ such that $\big||V_1|-|V_2|\big|\leq 1$ is possible. Hence
\begin{corollary}
If $Q\in\mathcal{Q}_3(n)$ has an independent set $I$ of size $|I| \geq 4n/10$ then it has a semi-equitable coloring of type 
$(|I|, \lceil (n-|I|)/2\rceil, \lfloor (n-|I|)/2 \rfloor)$.
\end{corollary}

Taking into account above considerations Theorem \ref{tw_res} can be extended as follows.

\begin{theorem}
If $Q\in\mathcal{Q}_3(n)$ and $\alpha(Q) \geq 4n/10$, then there exists an independent set $I$ of size $k$ in $Q$ such that $Q-I$ is bipartite for $\lfloor (n-\alpha(Q))/2 
\rfloor \leq k \leq \alpha(Q)$. \hfill $\Box$ \label{tw5}
\end{theorem}

The problem for $k < \lfloor (n-\alpha(Q))/2 \rfloor$ in $\mathcal{Q}_3(n)$ with $\alpha(Q) \geq 4n/10$ stays open as well as the 
problem for cubic graphs with $\alpha(Q) < 4n/10$.

Finally, note that Theorem \ref{tw5} cannot 
be generalized to all 3-colorable graphs, since the sun $S_3$ 
graph\footnote{$S_k$ consists of a central complete graph $K_k$ with an outer ring of k vertices, each of which is joined to both endpoints of the closest outer edge of the central core.} is a counterexample.

\vspace{0.5cm}
\noindent {\large \textbf{Acknowledgments}}

The authors thank Professor Darek Dereniowski for taking great care in reading our manuscript and making several useful suggestions improving the presentation.

\end{document}